\documentclass[fleqn,twoside]{article}
\usepackage{espcrc2}

\readRCS
$Id: espcrc2.tex,v 1.2 2004/02/24 11:22:11 spepping Exp $
\title{Landscape versus Swampland for Higher Derivative Gravity}

\author{Sho Yaida\address{Department of Physics, Stanford University, Stanford, CA 94305, USA}\thanks{The author was supported by the Albion Walter Hewlett Stanford Graduate Fellowship, the Stanford Institute for Theoretical Physics, and NSF Grant No. 0756174.}}

\begin{document}

\begin{abstract}
We survey recent studies of Gauss-Bonnet gravity and its dual conformal field theories, including their relation to the violation of the Kovtun-Starinets-Son viscosity bound.
Via holography, we can also study properties such as microcausality and unitarity of boundary field theory duals.
Such studies in turn supply constraints on bulk gravitational theories, consigning some of them to the swampland.
\end{abstract}

\maketitle

\section{LANDSCAPE AND SWAMPLAND}
String theory seems to possess a vast landscape of stable and metastable vacua~\cite{landscape}.
Some are grouped together by marginal deformations.
Some are related by dualities.
Tunneling processes connect isolated vacua.
Employing such processes, eternal inflation populates these vacua, and we may reside in one of the pocket universes which are well-described by Standard Model and General Relativity at sufficiently low energy.

By studying Calabi-Yau manifolds (with and without singularities), branes, fluxes, etc., we have obtained a partial map of the landscape and a long, yet incomplete, list of low energy effective field theories arising from string theory. One day, we may find our vacuum.

A question arises: given a randomly chosen effective field theory, can we always embed it into the string landscape? If not, we say that it belongs to the ``swampland." In this article, we review some attempts toward distinguishing theories in the swampland from those in the string landscape.

\section{GAUSS-BONNET GRAVITY}
As an example of seemingly innocuous low energy effective theories, let us take (4+1)-dimensional Gauss-Bonnet gravity with a negative cosmological constant $\Lambda \equiv-\frac{6}{L^2}$, described by the classical action of the form  (we suppress the associated Gibbons-Hawking surface term)
\begin{eqnarray}
\label{action}
I=\frac{1}{16\pi G_N} \mathop\int d^{5}x\sqrt{-g} \, [R-2\Lambda  \nonumber \\
  +
{\frac{\lambda_{GB}}{2}} L^2
(R^2-4R_{\mu\nu}R^{\mu\nu}+R_{\mu\nu\rho\sigma}R^{\mu\nu\rho\sigma})
].
\end{eqnarray}
Of course, in general, there is an infinite number of higher derivative terms. Here we are focusing on special cases where Gauss-Bonnet gravity dominates over the other higher derivative corrections.

Now the twist here is to use its holographic dual in order to show that, for certain values of $\lambda_{GB}$, it can never arise within the landscape.

Via the holographic dictionary, the above bulk gravitational theory defines a class of dual boundary conformal field theories (CFTs), parametrized by marginal deformations associated with varying $\lambda_{GB}$ in the bulk. There are various properties which we expect consistent field theories to satisfy, such as causality. If a theory does not satisfy such properties, we can consign it to the swampland.

\section{KSS VISCOSITY BOUND?}
Kovtun, Starinets, and Son~\cite{KSSbound} conjectured the following lower bound on the shear viscosity to entropy density ratio (KSS bound) for all consistent field theories:
\begin{equation}
\label{KSS}
\frac{\eta}{s}\geq\frac{1}{4\pi}.
\end{equation}
If true, we may use it to our advantage.

For example,~\cite{BLMSY1} and~\cite{CE1} have computed the shear viscosity to entropy density ratio for CFT duals of Gauss-Bonnet gravity. The finite $\lambda_{GB}$ calculation of~\cite{BLMSY1} gives
\begin{equation} \label{advertise}
\frac{\eta}{s}=\frac{1}{4\pi}[1- 4 \lambda_{GB}].
\end{equation}
If the KSS bound were true, then we could discard theories with positive $\lambda_{GB}$ as inconsistent.

However, nobody has proved the KSS bound. In fact,~\cite{CE1} and~\cite{CE2} construct consistent string theory backgrounds (and their dual CFTs) whose low energy descriptions involve Gauss-Bonnet gravity with $\Lambda<0$ and very small, yet positive, $\lambda_{GB}$. Therefore, they constitute counterexamples of the KSS bound (see also~\cite{CE3}). By studying Gauss-Bonnet gravity and its holographic duals, we have expanded our knowledge of field theoretic transport properties.

\section{MICROCAUSALITY}
Ordinary relativistic field theories satisfy microcausality: two local operators at spacelike separation (anti)commute. This in turn implies that retarded Green functions must vanish outside the lightcone. We expect dual CFTs to satisfy microcausality if they are consistent.

\cite{BLMSY2} argues that dual CFTs with $\lambda_{GB}>\frac{9}{100}$ have retarded Green functions that do not vanish outside the lightcone, thus violating microcausality. In this way, we conclude that Gauss-Bonnet gravity with negative cosmological constant and $\lambda_{GB}>\frac{9}{100}$ can never be embedded into the string landscape.

\section{STATUS REPORT}
The viscosity calculation and microcausality analysis of~\cite{BLMSY1,BLMSY2} have been extended to the Gauss-Bonnet theory with $U(1)$ charges in~\cite{Geetal} (see also~\cite{Caietal} for the case with dilaton). They again find the constraint $\lambda_{GB}\leq\frac{9}{100}$. They also see an interesting instability of charged black branes for $\lambda_{GB}>\frac{1}{24}$, but this does not necessarily render the theory inconsistent.

\cite{Hofman} investigates the implication of positive energy conditions. For supersymmetric CFTs, they find the following bound on $a$ and $c$ conformal anomaly coefficients: $\frac{3}{2}\geq\frac{a}{c}\geq\frac{1}{2}$. (They get a weaker bound for nonsupersymmetric CFTs.)
Applying this to CFT duals of Gauss-Bonnet gravity, we obtain the following constraint on $\lambda_{GB}$: $-\frac{7}{36}\leq\lambda_{GB}\leq\frac{9}{100}$. Interestingly enough, one end of the bound exactly coincides with the microcausality constraint. Role of supersymmetry here is mysterious.

Finally let us note that, although the KSS bound now seems to be violated, there may still be some universal lower bound on the shear viscosity to entropy density ratio. For example, with the microcausality constraint, CFTs with Gauss-Bonnet gravity dual do satisfy
\begin{equation}
\frac{\eta}{s}\geq\frac{16}{25}\frac{1}{4\pi}.
\end{equation}
Of course, we have established this bound only within a very small class of CFTs, and it does not have any right to be true in general. Nevertheless, it would be interesting if we could establish any nontrivial universal bound that applies to all consistent field theories.


\begin{thebibliography}{9}
\bibitem{landscape} For a review, see
  M.~R.~Douglas and S.~Kachru,
  Rev.\ Mod.\ Phys.\  {\bf 79}, 733 (2007)
  [arXiv:hep-th/0610102].
\bibitem{KSSbound}
  P.~Kovtun, D.~T.~Son and A.~O.~Starinets,
  Phys.\ Rev.\ Lett.\  {\bf 94}, 111601 (2005)
  [arXiv:hep-th/0405231].
\bibitem{BLMSY1} 
  M.~Brigante, H.~Liu, R.~C.~Myers, S.~Shenker and S.~Yaida,
  Phys.\ Rev.\  D {\bf 77}, 126006 (2008)
  [arXiv:0712.0805 [hep-th]].
\bibitem{CE1}
  Y.~Kats and P.~Petrov,
  JHEP {\bf 0901}, 044 (2009)
  [arXiv:0712.0743 [hep-th]].
\bibitem{CE2}
  A.~Buchel, R.~C.~Myers and A.~Sinha,
  arXiv:0812.2521 [hep-th].
\bibitem{CE3}
  A.~Adams, A.~Maloney, A.~Sinha and S.~E.~Vazquez,
  arXiv:0812.0166 [hep-th].
\bibitem{BLMSY2}
  M.~Brigante, H.~Liu, R.~C.~Myers, S.~Shenker and S.~Yaida,
  Phys.\ Rev.\ Lett.\  {\bf 100}, 191601 (2008)
  [arXiv:0802.3318 [hep-th]].
\bibitem{Geetal}
  X.~H.~Ge, Y.~Matsuo, F.~W.~Shu, S.~J.~Sin and T.~Tsukioka,
  JHEP {\bf 0810}, 009 (2008)
  [arXiv:0808.2354 [hep-th]].
\bibitem{Caietal}
  R.~G.~Cai, Z.~Y.~Nie, N.~Ohta and Y.~W.~Sun,
  arXiv:0901.1421 [hep-th].
\bibitem{Hofman}
  D.~M.~Hofman and J.~Maldacena,
  JHEP {\bf 0805}, 012 (2008)
  [arXiv:0803.1467 [hep-th]].
\end{thebibliography}
\end{document}